\documentstyle[12pt]{article}

\newcommand{\refpa}[1]{(\ref{#1})}
\newcommand{\be}{\begin{equation}}
\newcommand{\ee}{\end{equation}}
\newcommand{\bea}{\begin{eqnarray}}
\newcommand{\eea}{\end{eqnarray}}

\begin{document}
\begin{titlepage}
\begin{flushleft}
G\"oteborg\\
ITP 94-19\\
hep-th/9407035\\
June 1994\\
\end{flushleft}
\vspace{1cm}
\begin{center}
{\Large SL(2,{\bf R}) YANG-MILLS THEORY ON A CIRCLE}\\
\vspace{5mm}
{\large Ingemar Bengtsson}\footnote{Email address: ingemar@vana.physto.se}\\
{\sl Fysikum\\
University of Stockholm\\
Box 6730, S-113 85 Stockholm, Sweden}\\
\vspace{1cm}
{\large Joakim Hallin}\footnote{Email address: tfejh@fy.chalmers.se}\\
{\sl Institute of Theoretical Physics\\
Chalmers University of Technology\\
and University of G\"oteborg\\
S-412 96 G\"oteborg, Sweden}\\
\vspace{1.5cm}
{\bf Abstract}\\
\end{center}
The kinematics of SL(2,{\bf R}) Yang-Mills theory on a circle is considered,
for reasons
that are spelled out. The gauge transformations exhibit hyperbolic
fixed points, and this results in a physical configuration space with a
non-Hausdorff "network" topology. The ambiguity encountered in canonical
quantization is then much more pronounced than in the compact case, and can
not be resolved through the kind of appeal made to group theory in that case.
\end{titlepage}

\noindent We have studied Yang-Mills theory on a cylindrical space-time,
choosing the non-compact group SL(2,{\bf R}) for our structure group. Since
this undertaking may appear peculiar, we will begin by spelling out our
motivation. First of all the Yang-Mills Hamiltonian is not positive definite
whenever the structure group is non-compact. However, this is of no concern to
us, since this operator will have nothing to do with the time-development of
our model. Actually, we will not be concerned with time-development at all, so
that we are really interested only in setting up the model on a space which has
the topology of the circle - ``Yang-Mills theory'', here, refers only to the
phase space of the model. Our interest has to do with gravity. We know that
there are four choices of structure group for Yang-Mills theory that are of
physical interest: U(1), SU(2) and SU(3) - all of which are compact - and
SL(2,{\bf C}), which is non-compact. The latter case can be used to formulate
Einstein's theory \cite{Abhay1}. We believe that it is important to gain a
broad experience of non-compact gauge theories, and this is one of two reasons
for studying the toy model that we will describe. (It is true that the group
theory of SL(2,{\bf R}) differs in important ways from the group theory of
SL(2,{\bf C}), but the real case has certain simplifying features, and it
seemed worthwhile to do a separate study of this case.)

This is our first motivation. The second motivation is more vague, but at least
as important. So far, almost all our intuition about gauge theories comes from
Yang-Mills theory with compact structure groups. If we step back a bit from the
problem, and view the gauge transformations in the same way as we might view
the Hamiltonian flow of a dynamical system, we observe that we are then dealing
with gauge transformations of a very simple, ``integrable'' kind. Presumably,
this is not a generic case, and presumably the ``gauge flow'' in any theory of
gravitation - where time development itself may be viewed as a gauge
transformation - is a very different kettle of fish. We believe that there is a
risk that our experience from compact Yang-Mills theory may be qualitatively
misleading when it comes to defining a quantum theory of gravity. Our toy model
is of some interest here - although the gauge flow remains integrable, it
exhibits hyperbolic fixed points, which is at least a step in the ``chaotic''
direction.

We hope that we have convinced the reader that we have chosen an interesting
subject, and we will now briefly review the properties of Yang-Mills theory on
a circle, and how one may set up the quantum theory in the compact case. To
make a long story \cite{Rajeev1} short, the physical configuration space of the
model is the space of conjugacy classes in the group. This is the basic fact on
which our analysis is based. Hence we have a finite dimensional problem in
front of us. Nevertheless, as readers familiar with the original references
will recognize, there is enough structure left to make the toy model
interestingly analogous to the 3+1 dimensional theory. Let us draw attention to
 some 1+1 dimensional peculiarities that are of particular interest to us. In
the spirit of the ``loop quantization program'' \cite{Lee1}, we wish to use
traces of holonomies, with inserted momenta, as coordinates on the physical
phase space. So we define

\begin{equation} T^0(n) = \mbox{tr}\, h(x)^n \hspace*{2cm} T^1(x,n) = \mbox{tr}
\,E(x)h(x)^n , \end{equation}

\noindent where $h(x)$ is the holonomy. We also define an object which we will
call the Hamiltonian, although we will not use it to generate time-evolution:

\begin{equation} H = -\frac{1}{2} \mbox{tr} \,E(x)^2 . \end{equation}

\noindent It is then a peculiarity of 1+1 dimensions that

\begin{equation} \{T^0(n),H\} = 4inT^1(n) . \end{equation}

\noindent In other words, $H$ can be used to generate the higher $T$-variables
\cite{Jocke1}. Therefore it suffices to consider $T^0$ and $H$. A 1+1
dimensional
peculiarity which is of some interest to us is that the generator of spatial
diffeomorphisms is related to the generator of gauge transformations by

\begin{equation} {\cal D}(x) = \mbox{tr}\,A(x){\cal G}(x) , \end{equation}

\noindent where $A$ is the connection, a spatial scalar density. This means
that a gauge invariant object is also diffeomorphism invariant, and it follows
that the Hamiltonian, as defined above, is weakly x-independent.

Now what we want to do is to obtain an explicit description of the physical
configuration space for specific choices of the structure groups, and then to
set up a quantum version of the theory, in which the wave functions are
functions of the physical configuration space and the operators $T^0$ and $H$
are
realized as self-adjoint operators. Let us first recall how this goes for
compact structure groups \cite{Rajeev1}, choosing SU(2) as our example. In this
case the group manifold is $S^3$, the three-sphere. The unit and the anti-unit
elements - which we can imagine as sitting at the South and North Pole of the
sphere, respectively, form conjugacy classes by themselves.

\begin{figure}[h]
\begin{picture}(300,100)(-90,0)
\put(100,50){\circle{55}}
\put(200,25){\line(0,1){50}}
\end{picture}
\caption{SU(2) and its conjugacy classes}
\label{sfar}
\end{figure}
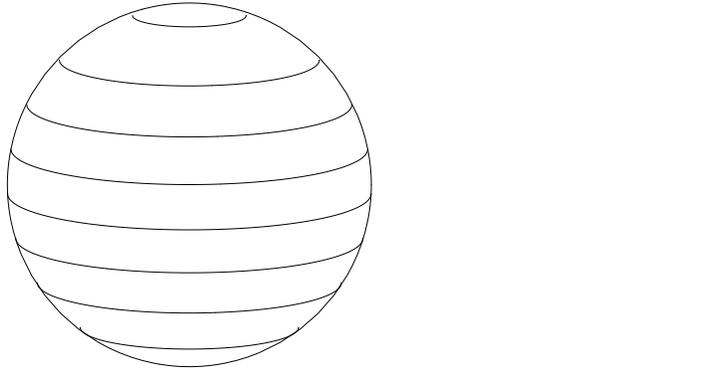

A line of constant longitude between the poles is a Cartan subgroup, and the
conjugacy class to which any point in the Cartan subgroup belongs is given by
all the points on the three-sphere that have the same latitude as the given
points. In this way, the group manifold is nicely foliated by the conjugacy
classes, and the space of conjugacy classes is simply a closed interval. For
some purposes (notably for the generalisation to arbitrary compact groups), it
is more suitable to define this interval as $S^1/{\pi}^2$, where the
permutation group ${\pi}^2$ is in fact the Weyl group, acting on the circle in
a suitable way.
Since the action of the discrete group has fixed points, this is an orbifold
rather than a manifold. From fig. \ref{sfar}\footnote{The figures are
incomplete. A complete version will be supplied to the
appropriate journal.}, it is clear that the orbifold singularities appear
because there are elliptic fixed points in the gauge flow. This picture
generalizes in a straightforward manner to arbitrary compact structure groups.
Moreover, the essential feature that the physical configuration space is an
orbifold generalizes to the 3+1 dimensional case \cite{Mitter1}.

After canonical quantization \cite{Rajeev1}, the operators to be made
self-adjoint become

\begin{equation} T^0(n) = 2\cos{n{\phi}} \hspace*{2cm} H = - \partial^2
_{\phi}.
\label{SU(2)}\end{equation}

\noindent $T^0$ causes no problems, but $H$ does not have a unique self-adjoint
extension. To make it self-adjoint, we may specify Dirichlet or Neumann
conditions, or any combination of those, at the ends of the interval. The
choice will affect the spectrum of $H$. If one wants a unique, or at least a
preferred, answer one has to add further rules to the game. We might insist
that the result should be the same if we choose to quantize before constraining
the theory, or that the result should be in some sense stable against addition
of matter degrees of freedom to the model. In the present case, a preferred
choice also emerges after an appeal to group theory \cite{Rajeev1}. After a
canonical transformation of the operators given in eq. (\ref{SU(2)}) one can
choose the boundary conditions such that the Hilbert space measure becomes the
Haar measure restricted to conjugacy classes, the Hamiltonian becomes the
restriction of the Laplacian defined on the group manifold, and its eigenstates
become the characters of SU(2). (The interplay between the canonical
transformation and the Weyl group gives rise to a subtlety here, which does not
arise in the SL(2,{\bf R}) case. We refer to the literature for this
\cite{Jocke1}.) This choice is clearly in a sense to be preferred.

\begin{figure}[b]
\begin{picture}(300,120)(-95,0)
\put(10,10){\framebox(50,100)}
\put(35,10){\line(1,1){25}}
\put(35,10){\line(-1,1){25}}
\put(10,35){\line(1,1){50}}
\put(60,35){\line(-1,1){50}}
\put(35,110){\line(1,-1){25}}
\put(35,110){\line(-1,-1){25}}
\put(200,50){\circle{50}}
\put(221,50){\line(1,0){30}}
\put(147,50){\line(1,0){30}}
\put(180.3,50){\circle*{4}}
\put(219.8,50){\circle*{4}}
\put(70,7){{\tiny $\theta = 0$}}
\put(70,57){{\tiny ${\theta} = {\pi}$}}
\put(70,107){{\tiny ${\theta} = 2{\pi}$}}
\put(27,5){{\tiny $r = 0$}}
\end{picture}
\caption{SL(2,{\bf R}) and its conjugacy classes}
\label{hogvatten}
\end{figure}
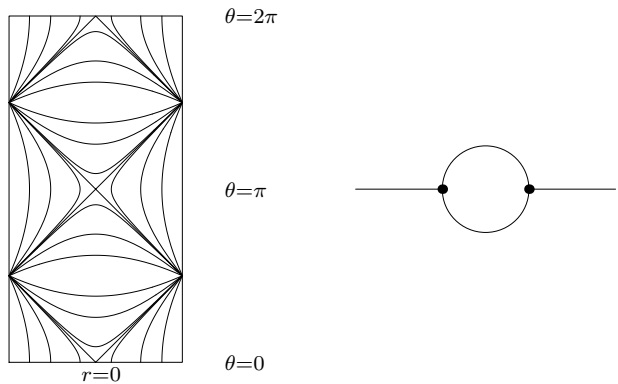

Let us now see how things change when we move on to the non-compact structure
group SL(2,{\bf R}). The group manifold is now three dimensional anti-de Sitter
space, which we can depict as the Penrose diagram
drawn in fig. \ref{hogvatten}. We also use the G = KAN decomposition of the
group, which we
can write in terms of matrices as

\begin{eqnarray} \left(\begin{array}{ll} Y + Z & X + T
\\ X - T & - Y + Z \end{array}\right) =
\left( \begin{array}{ll} \cos{{\theta}} & \sin{{\theta}}
\\ - \sin{{\theta}} & \cos{{\theta}} \end{array}\right)
\left( \begin{array}{ll} e^t & 0
\\ 0 & e^{-t} \end{array}\right)
\left( \begin{array}{ll} 1 & s
\\ 0 & 1 \end{array}\right) ; \nonumber \\
 - X^2 - Y^2 + Z^2 + T^2 = 1. \hspace*{3cm} \end{eqnarray}

\noindent In the Penrose diagram, the compact subgroup K corresponds to the
line $r = 0$, the hyperbolic subgroup A corresponds to the line ${\theta} = 0$,
and the
parabolic subgroup N corresponds to the lightcone with vertex at the origin,
i.e. at the unit element. We note that

\begin{equation} TrG = 2Z \hspace*{1cm} TrK = 2\cos{\theta} \hspace*{1cm} TrA =
2\cosh{t} \hspace*{1cm} TrN = 2 . \end{equation}

\noindent It is then easy to deduce how the group is foliated by the conjugacy
classes. We need only one additional piece of information, which is that the
parameter s in N may be scaled to plus or minus one, but it can not be set
equal to zero by conjugation. Hence the unit and anti-unit element form
conjugacy classes by themselves, and the backwards and forwards light cones
from these points also form conjugacy classes. The conjugacy class that
contains a given element of K (not equal to the unit or anti-unit elements)
forms one sheet of a two-sheeted hyperboloid lying within these light cones,
and the conjugacy class that contains a given element of A forms a one-sheeted
hyperboloid surrounding the same light cones. This is depicted in fig.
\ref{hogvatten}. It is intuitively obvious what the topology of the space of
conjugacy classes is, given the topology of the group, and this has also been
drawn in the figure. The resulting topology exhibits three features that are
new compared to the SU(2) case: It is non-compact, it is non-Hausdorff, and it
has the structure of a ``network'' rather than that of a manifold of a fixed
dimension.

We observe that the picture can be generalized to SL(N, {\bf R}). Of more
importance is the fact that the topology of the physical configuration space of
non-compact Yang-Mills theory is known to be non-Hausdorff also in 3+1
dimensions \cite{Abhay2}. Hence our toy model captures some aspects of the
physically interesting case.

Let us comment on the non-Hausdorff property. The enlarged dots in the figure
each
denote three separate points, corresponding to the backwards and forwards light
cone and its vertex, respectively. These points can not be separated by any
continuous function. It is geometrically clear how this complication arises
when we take the quotient of the group manifold by conjugations (indeed
non-Hausdorff spaces typically arise in some such way, when they arise at all).
It is also clear that this is a harmless complication for many purposes,
especially when we go on to consider quantum mechanics on the space of
conjugacy classes. The wave function at a point does not matter. Hence the
non-Hausdorff property can be safely ignored in the sequel.

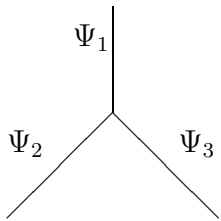
\begin{figure}[b]
\begin{picture}(200,100)(-135,0)
\put(100,50){\line(0,1){40}}
\put(100,50){\line(1,-1){40}}
\put(100,50){\line(-1,-1){40}}
\put(85,75){${\Psi}_1$}
\put(60,35){${\Psi}_2$}
\put(125,35){${\Psi}_3$}
\end{picture}
\caption{Three half-lines meeting at a point.}
\label{vertex}
\end{figure}

The network structure is important - it will be necessary to supply appropriate
boundary conditions at the vertices in order to ensure that the Hamiltonian be
self-adjoint. Quantum theory on networks has been considered in quantum
chemistry \cite{Ruedenberg} and more recently in connection with mesoscopic
networks \cite{Bal1} - in which case the networks are typically made out of
gold films, say ten nanometers thick. Let us review the simplest case of three
half-lines meeting at a point, as in fig. \ref{vertex}. The Hilbert space is
$L^2$ of the union of the three halflines $[0,\infty [$, and we choose the
Hamiltonian to be the flat Laplacian on each half-line. This is a symmetric
operator, and it follows from general theory \cite{Reed} that it  admits a nine
parameter family of self-adjoint extensions (while the translation operator
admits none). The domain of definition of a self-adjoint extension is given
explicitly \cite{Exner1} by the three conditions

\begin{equation} \Psi '_i(V) = K_{ij}{\Psi}_j(V) , \label{K} \end{equation}

\noindent where $K_{ij}$ is a hermitian matrix. In applications to solid state
physics, the network is only an approximation to a network of finite thickness.
Physical intuition dictates that the wave function should be continuous at the
vertex (possibly up to a phase). Then there is only a one-parameter family of
self-adjoint extensions left, defined by

\begin{equation} \Psi '_1(V) + \Psi '_2(V) + \Psi '_3(V) = {\lambda}{\Psi}_1(V)
= {\lambda}{\Psi}_2(V) = {\lambda}{\Psi}_3(V) ; \hspace*{5mm} {\lambda} \in
{\bf R} . \label{solidstate} \end{equation}

\noindent The condition on the derivatives enforces conservation of the
probability current, and a non-zero value of the parameter ${\lambda}$ can be
thought of as a delta function potential at the vertex. When setting boundary
conditions for an entire network, the vertices are treated separately.

\begin{figure}[t]
\begin{picture}(300,110)(-30,0)
\put(200,50){\circle{90}}
\put(120,50){\line(1,0){60}}
\put(220,50){\line(1,0){50}}
\put(190,80){$\Psi_1(\theta)$}
\put(190,10){$\Psi_2(\theta)$}
\put(130,60){$\Psi_4(t)$}
\put(250,60){$\Psi_3(t)$}
\put(250,35){$I$}
\put(130,35){$II$}
\end{picture}
\caption{The SL(2,{\bf R}) network.}
\label{network}
\end{figure}
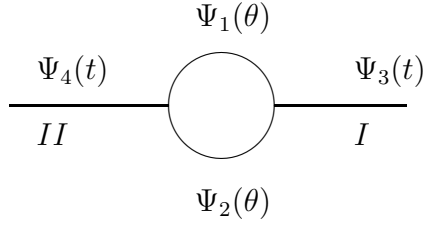

Let us now turn to our network, which is the space of conjugacy classes of
SL(2,{\bf R}). We begin by defining the coordinates that we will use, as well
as the form of the Hamiltonian operator on the various segments. This is done
in fig. \ref{network}. The $T^0$-variable is self-adjoint as it stands. The
fact that the Hamiltonian is no longer positive definite causes no particular
problems - for a single vertex, all that happens is that the matrix $K_{ij}$ in
eq. (\ref{K}) becomes pseudo-hermitian. The deficiency indices \cite{Reed} of
the Hamiltonian on the full network are (6,6), so there is a 36-parameter
family of self-adjoint extensions of this operator. Unless we add further
requirements, all of them are in a sense ``correct''. It does seem reasonable
to require that the conditions should be set at each vertex separately, but a
large ambiguity remains. Clearly, physical intuition is not necessarily a good
guide for setting boundary conditions here. In the compact case, we saw that an
appeal to group theory was enough to single out a preferred answer. Now the
qualitative properties of the spectrum of $H$ will be the same for most
 choices. There will be a discrete set of levels bounded from below, and a
 doubly
degenerate continuous set bounded from above. This is reasonable from the point
of view of group theory, and corresponds very roughly speaking to the
discrete and
principal series of representations, respectively (the supplementary series
plays no r\^{o}le in harmonic analysis). Important qualitative issues are
nevertheless at stake, in particular whether superselection rules will occur.
We could impose (say) Dirichlet conditions on all ends, in which case they
certainly do. On the other hand, with the ``solid state'' conditions in eq.
(\ref{solidstate}), there are no superselection rules: The wave function will
leak through the vertex when $H$ is applied to it.

Let us begin the discussion by imposing the solid state conditions at each
vertex. This will provide us with an explicit example of a quantum version of
our model. More precisely, taking the various signs into account and using the
coordinates given in fig. \ref{network}, we impose

\begin{eqnarray}
\Psi '_1(0) - \Psi '_2(2{\pi}) - \Psi '_3(0) &=& {\lambda}{\Psi}_1(0) =
{\lambda}{\Psi}_2(2{\pi}) =
{\lambda}{\Psi}_3(0) \\
- \Psi '_1(\pi )+\Psi '_2(\pi )-\Psi '_4(0) &=& {\lambda}{\Psi}_1(\pi ) =
{\lambda}{\Psi}_2(\pi ) =
{\lambda}{\Psi}_4(0)
\end{eqnarray}
\noindent where $\lambda$ is a real parameter. The Hamiltonian H is given by
$-\partial _{\theta}^2$ on the circle and $\partial _t^2$ on the half-lines.
Note also that $T^0(n)$ is given by $2\cos (n\theta )$ on the circle,
$2\cosh (n t)$ on half-line $I$ and $-2\cosh (n t)$ on half-line $II$.
The eigenvalues $E$ of $H$ are

\[ E = k^2 \]

\noindent for suitable values of $k$. The possible values of $k$ are as
follows.
There is a discrete set of states with support on the entire network and
oscillatory behaviour on the circle, for which
\begin{equation}
e^{2\pi i k}=\left( \frac{1-\lambda /k+2i}{1-\lambda /k-2i}\right) ^2 ,\; \;
\; k>0.
\end{equation}
For all values of ${\lambda}$
there will be an additional set of discrete states with support confined to the
circle, and

\begin{equation} k = n \in {\bf Z_+} . \end{equation}
\noindent Precisely when ${\lambda}$ is a positive integer, there are two
additional states, smooth on
the circle and with support on the entire network, for which
\be
k=\lambda =n.
\ee
\noindent A state with zero eigenvalue of $H$ exists only when

\begin{equation} {\lambda} = 0 \hspace*{3mm} or \hspace*{3mm} {\lambda} = -
\frac{4}{{\pi}} . \end{equation}

\noindent For all values of ${\lambda}$ there is a doubly degenerate set of
states with oscillatory behaviour on the half-lines and with continuous
negative eigenvalues of $H$.

Finally, we discuss whether an appeal to group theory will help to cut down the
ambiguity that we have encountered. (A good reference for harmonic analysis on
SL(2,{\bf R}) is the book by Varadarajan \cite{Varadarajan}.) The first
observation is that the characters of SL(2,{\bf R}) do not obey boundary
conditions
that satisfy \refpa{K}. The characters of the discrete series are

\begin{equation} {\Theta}_n(\theta ) = -
\mbox{sgn}(n)\frac{e^{in{\theta}}}{e^{i{\theta}} - e^{-i{\theta}}}
\label{character1} \end{equation}

\noindent on the circle,

\begin{equation} {\Theta}_n(t) = \frac{e^{-|n| t}}{e^t-e^{-t}},\end{equation}
on half line I and
\be
\Theta _n (t)=(-1)^{n-1} \frac{e^{-|n| t}}{e^t-e^{-t}}, \label{character3}
\ee
on half line II. The next, trivial but important, observation is that a class
function can not be square integrable on the group, when the latter is
non-compact. What we can do is to associate a class function with any smooth
function $f(g)$ of compact support on the group, by means of an orbital
integral. Let B denote the elliptic Cartan subgroup (which corresponds to
the circle), and L denote the hyperbolic Cartan subgroup (the half-lines).
Then the orbital integral associated to B is

\begin{equation} F_{f,B}(\theta ) = (e^{i\theta }-e^{-i\theta }) \int_{G/B}
f(gu_{\theta}g^{-1})d\dot{g} , \end{equation}

\noindent where $u_{\theta}$ is an element of B and $d\dot{g}$ is an
invariant measure on G/B. There is a similar formula for L. The
normalizing factor in front of the integral has been chosen so that the
Laplacian
${\Omega}$ on the group gets pushed down precisely to our network
Hamiltonian:

\begin{eqnarray} F_{{\Omega}f,B} = - \frac{d^2}{d\theta ^2} F_{f,B} = H F_{f,B}
\\
F_{{\Omega}f,L} = \frac{d^2}{dt^2} F_{f,L} = H F_{f,L} .  \end{eqnarray}

\noindent It is now crucial to determine the behaviour of the orbital integrals
at the network vertices. One finds e.g.

\begin{equation} \frac{1}{i}(F_{f,B}(0^+) - F_{f,B}(0^-)) = F_{f,LI}(0) .
\end{equation}

\noindent Also the first derivative of $F_{f,B}$ is continuous on B, and
$F_{f,L}$ obeys the Neumann condition. The discontinuity of $F_{f,B}$ is called
the Harish-Chandra jump relation, and - apart from
factors which depend on the specific normalization of the orbital integrals
used - it is intuitively clear why it occurs, since the integral over
 the one-sheeted hyperboloids in anti-de Sitter space
will tend to the sum of the integrals over the two sheets of the two-sheeted
hyperboloid as both surfaces approach the light cone. Unfortunately, for our
purposes, this is not an acceptable behaviour for a wave function, since these
boundary condition do not give the domain of definition of a self-adjoint
Hamiltonian.

The objects that are naturally integrated against the orbital integrals are
characters (times a normalizing factor, so that the denominators in eqs.
(\ref{character1} - \ref{character3}) is removed), which obey e.g.
\be
-i \Phi '_B(0)=\Phi '_{LI}(0).
\ee

\noindent Also ${\Phi}_B$ is a smooth function on B. These are the matching
conditions of Harish-Chandra. Then integration by parts can be done in

\begin{equation} \int {\Theta}fdG = -\int _{0}^{2\pi}{\Phi}_BF_{f,B}d{\theta} +
\int_{0}^{\infty} {\Phi}_{LI}F_{f,LI}dt +
\int_{0}^{\infty}{\Phi}_{LII}F_{f,LII} dt , \end{equation}

\noindent with no boundary terms. Unfortunately, this is not useful for our
purposes.

In conclusion, we have investigated the physical configuration space of SL(2,
{\bf R}) Yang-Mills theory in 1+1 dimensions. The ``gauge flow'' exhibits
hyperbolic fixed points, which leads to topological complications that are not
present for compact structure groups, and results in a considerable ambiguity
in the quantum theory. Unlike the ambiguity that arises in the compact case,
this ambiguity affects broad qualitative issues such as the appearance of
superselection rules, and we were not able to resolve it through any
straightforward appeal to group theory. A further study seems called for - as a
natural second step, one could consider BRST quantization.

\vspace*{2cm}

{\em Acknowledgements:}

\vspace*{5mm}

\noindent We thank those of our friends who listened. Special thanks to Ralph
Howard.

\end{document}